\documentclass[prb,showpacs,floatfix,superscriptaddress,preprint]{revtex4-1}

\usepackage{graphicx}

\usepackage{color}

\begin{document}
%
%
\title{Ab initio lattice dynamics and electron-phonon coupling of Bi(111)}

\author{M.~Alc\'{a}ntara Ortigoza}
\affiliation{Department of Physics,
University of Central Florida
Orlando, Florida 32816, USA}
\affiliation{Institut f\"ur Festk\"orperphysik,
             Karlsruher Institut f\"ur Technologie,
             D-76021 Karlsruhe, Germany}

\author{I.~Yu. Sklyadneva}
\affiliation{Donostia International Physics Center (DIPC),
20018 San Sebasti\'an/Donostia, Basque Country, Spain}
\affiliation{Institut f\"ur Festk\"orperphysik,
             Karlsruher Institut f\"ur Technologie,
             D-76021 Karlsruhe, Germany}
\affiliation{Tomsk State University,6340501, Tomsk, Russian
Federation}

\author{R.~Heid\cite{corr}}
\affiliation{Institut f\"ur Festk\"orperphysik,
             Karlsruher Institut f\"ur Technologie,
             D-76021 Karlsruhe, Germany}

\author{E.~V.~Chulkov}
\affiliation{Donostia International Physics Center (DIPC),
20018 San Sebasti\'an/Donostia, Basque Country, Spain}
\affiliation{Departamento de F\'{i}sica de Materiales, Facultad de
Ciencias Qu\'{\i}micas, UPV/EHU, Apdo. 1072, 20080 San
Sebasti\'an/Donostia, Basque Country, Spain}
\affiliation{Tomsk State University,6340501, Tomsk, Russian
Federation}
\affiliation{Centro
de F\'isica de Materiales CFM - Materials Physics Center MPC,
Centro Mixto CSIC-UPV/EHU, 20018 San Sebastian/Donostia, Spain}

\author{T.~S.~Rahman}
\affiliation{Department of Physics,
University of Central Florida
Orlando, Florida 32816, USA}

\author{K.-P.~Bohnen}
\affiliation{Institut f\"ur Festk\"orperphysik,
             Karlsruher Institut f\"ur Technologie,
             D-76021 Karlsruhe, Germany}

    \author{P.~M.~Echenique}
\affiliation{Donostia International Physics Center (DIPC), 20018
San Sebasti\'an/Donostia, Basque Country, Spain}
\affiliation{Departamento de F\'{i}sica de Materiales, Facultad de
Ciencias Qu\'{\i}micas, UPV/EHU, Apdo. 1072, 20080 San
Sebasti\'an/Donostia, Basque Country, Spain} 
\affiliation{Centro
de F\'isica de Materiales CFM - Materials Physics Center MPC,
Centro Mixto CSIC-UPV/EHU, 20018 San Sebastian/Donostia, Spain}

\date{Version of \today}

\begin{abstract}
We present a comprehensive {\it ab initio} study of structural,
electronic, lattice dynamical and electron-phonon coupling
properties of the Bi(111) surface within density functional
perturbation theory. Relativistic corrections due to spin-orbit
coupling are consistently taken into account. As calculations are
carried out in a periodic slab geometry, special attention is given
to the convergence with respect to the slab thickness. Although the
electronic structure of Bi(111) thin films varies significantly with
thickness, we found that the lattice dynamics of Bi(111) is quite
robust and appears converged already for slabs as thin as 6
bilayers. Changes of interatomic couplings are confined mostly to
the first two bilayers, resulting in super-bulk modes with
frequencies higher than the optic bulk spectrum, and in an enhanced
density of states at lower frequencies for atoms in the first
bilayer. Electronic states of the surface band related to the outer part of the hole
Fermi surfaces exhibit a moderate electron-phonon coupling of about
0.45, which is larger than the coupling constant of bulk Bi. States
at the inner part of the hole surface as well as those
forming the electron pocket close to the zone center show much
increased couplings due to transitions into bulk
projected states near $\overline{\Gamma}$. For these cases,
the state dependent Eliashberg functions exhibit pronounced peaks
at low energy and strongly deviate in shape from a Debye-like
spectrum, indicating that an extraction of the coupling strength
from measured electronic self-energies based on this simple
model is likely to fail.
\end{abstract}


\maketitle

\section{Introduction}

Looking back since the advent of modern solid state physics, bismuth
(Bi) has been perhaps one of the most intriguing element for research.
Its singular properties make Bi the metal with the lowest thermal
conductivity after mercury and the element with largest diamagnetism. Bi
allowed an early discovery of the Seebeck, the de Haas-van Alphen, the
Shubnikov-de Haas, and the Nernst effects, all of which are inherently
present in metals but were more challenging to observe in general. Bismuth
was in fact the first metal whose Fermi surface was experimentally
identified \cite{shoenberg39} and provided the basis to determine that of
other metals. Moreover, bulk Bi was the first non-superconducting material
(at least in the ordered phase) that was found to be superconducting in
structurally bulk-like nanoparticles. \cite{weitzel}

Many of the outstanding properties of bulk Bi are linked to its peculiar
electronic structure.  Bi is a semimetal with tiny electron and hole
pockets and a very small density of states (DOS) at the Fermi level.
Metallic screening is therefore much weaker than in typical metals,
and the interatomic bonding has a stronger directional component. This
favors a layered crystal structure with alternating weak and strong
interlayer bonds, which can be described as a stacking of bilayers.

Being a semimetal with low DOS at the Fermi energy,
bulk Bi is expected to have a small average electron-phonon ($e$-ph) coupling
constant, $\lambda$. Simple estimates based on a tight-binding
model\cite{gayone03} as well as a very recent {\it ab initio}
calculation\cite{huang13} yielded rather small values of $\lambda$=0.13
and 0.09, respectively, consistent with the finding that Bi does
not display superconductivity down to 50 mK.\cite{tian08} However,
since long ago amorphous bulk Bi was found to have a quite strong
electron-phonon coupling of $\lambda$=2.46 (the value is larger than
that of amorphous lead, \cite{chen71} which is known to have a strong
electron-phonon coupling also in the crystalline form \cite{heid10}).
This indicates that the intrinsic coupling strength of electronic states
to the phonons is comparable to those of lead, but superconductivity is
only absent due to the very low DOS at $E_F$.

Interest in surface properties of Bi arises on the one hand from the
large difference of the surface electronic structure from the bulk
one. Typically, an enhanced metalicity at the surface is observed
which is related to the appearance of surface electronic states
with an increased DOS at the Fermi energy ($E_F$).\cite{hofmann06}
On the other hand, spin-orbit interaction is strong for Bi, and is a
necessary ingredient for even a qualitative description of the electronic
structure already for the bulk. On surfaces it invokes large splittings of surface states with
deep implications for the Fermi surface topology. Among the low-index
surfaces, Bi(111) is a kind of model surface for investigating these
properties, because it does not break the bilayer structure. This enables
the study of modifications in the electronic structure originating in
the loss of translational symmetry in a pure form without complications
due to additional breaking of chemical bonds.  The surface electronic
structure of Bi(111) has been elucidated in a series of photoemission
experiments,\cite{jezequel86,hengsberger00,ast01,ast02,ast04,koroteev04}
which established the enhanced density of charge carriers and the presence
of surface states split by spin-orbit interaction.  These states create
two types of Fermi surfaces, known as hole and electron pockets.

The enhanced DOS at Bi surfaces has stimulated experimental investigations
of the coupling strength of surface electronic states.  For the Bi(111)
surface Ast and H\"{o}chst have estimated the electron-phonon coupling
from spectral functions of angular-resolved photoemission spectroscopy
(ARPES) measurements assuming a Debye model to account for the
spectral shape.\cite{ast02} The results turned out to be
strongly dependent on the cutoff frequency of the assumed Debye
spectrum. Fitting the
data with a cutoff Debye frequency of 10~meV, appropriate for bulk Bi,
yields $\lambda$=0.6, while assuming a surface Debye frequency of 5 meV
yields $\lambda$=2.3. This strong difference was later ascribed mainly to
the limited accuracy of the experiment.\cite{kirkegaard05} An alternative
road was pursued by Gayone {\em et al.}, in which the imaginary part of
the self-energy was extracted from the temperature dependence of the
linewidth of momentum distribution curves.\cite{gayone05} They found
smaller values of $\lambda$=0.40(5) for the surface electronic states,
albeit still significantly larger than the bulk value.

A prerequisite for a detailed analysis of the electron-phonon coupling
at surfaces is the knowledge of the surface vibrational spectrum. In
particular, surface localized vibrational modes can couple more strongly
to surface electronic states due to their larger spatial
overlap.\cite{hofmann09}  In thin films, it has been shown
that important contributions to the coupling may come from optical
vibrations.\cite{sklyadneva11} Experimentally the surface phonon spectrum
of Bi(111) has been investigated very recently via inelastic Helium atom
scattering (HAS) by Tamt\"{o}gl {\em et al.}\cite{tamtoegl10,tamtoegl13}
They found a rich spectrum of localized modes, including super-bulk
modes with frequencies above the maximum bulk frequency.

In view of the experimental uncertainties in extracting reliable numbers
for the electron-phonon coupling, theoretical calculations on an ab
initio basis are highly desirable. The density functional perturbation
theory (DFPT) provides a unified scheme to predict electronic, phonon
and electron-phonon coupling properties.  However, applications of this
technique to bulk Bi and in particular to Bi surfaces are still
a challenge, because the inclusion of spin-orbit interaction
increases the computational effort substantially, especially for
calculations of lattice dynamical quantities. Therefore, work along this
line was first devoted to the electronic structure. 
Large splittings of surface bands due to spin-orbit interaction were 
found for all low-index Bi surfaces by Koroteev {\em et al.}\cite{koroteev04} 
In a later work, it was shown that the surface electronic structure of
Bi(111) converges very slowly with increasing thickness of the slab, in
particular near the zone boundary at $\overline{\rm M}$.\cite{koroteev08}

Applications of DFPT to lattice dynamical properties including
spin-orbit coupling (SOC) have been devoted first to bulk
Bi.\cite{murray07,diaz07} D\'{i}az-S\'{a}nchez {\em et al.} showed
that inclusion of SOC is crucial for an improved agreement between
calculation and measurement,\cite{diaz07} but it did not provide such
a good description as it was the case, for example, for the phonon
dispersion of lead.\cite{heid10}. Very recently, DFPT has been used
to investigate the lattice dynamics of Bi(111). Due to the numerical
effort to include spin-orbit interaction, only slabs up to 6 bilayers
could be treated,\cite{tamtoegl13,chis13,huang13,yang13,benedek14},
raising the question to what extent the obtained results represent
the surface dynamics in the limit of macroscopically thick slabs. For
the 6-bilayer calculation, good agreement with the HAS data was
found.\cite{tamtoegl13,benedek14}

No attempt has been made so far to address the question of electron-phonon
coupling of surface states within such an {\it ab initio} approach.  The aim
of this paper is to provide a comprehensive {\it ab initio} analysis of the
electronic structure, lattice dynamics, and electron-phonon coupling of
surface electronic states for the Bi(111) surface.  Towards this goal, we
will also discuss the bulk lattice dynamics, which enters the evaluation
of the surface vibrations.  As we want to focus on the properties of a
semi-infinite surface, we carefully investigate the convergence of the
various quantities with increasing thickness of the slabs.

The paper is organized as follows.
The computational details are described in the following Section.
Results are presented and discussed in Sec.~\ref{sec_results}.  First we
briefly recapitulate the structure and lattice dynamics of bulk Bi in
Sec.~\ref{subs_bulk} which is a prerequisite for the surface dynamics
study, and present results for the electron-phonon coupling constant.
Secs.~\ref{subs_111geo} to \ref{subs_111pho} are devoted to the
structural, electronic, and lattice dynamical properties of the Bi(111)
surface, respectively, with emphasis on convergence with respect to the
slab thickness.  The coupling of surface localized electronic states is
then discussed in Sec.~\ref{subs_epc}.  Finally the results are summarized
in Sec.~\ref{sec_sum}.


\section{Computational details}
\label{sec_comp}

We performed density functional theory (DFT) calculations of bulk Bi and
Bi(111) within the local density approximation (LDA) in the parameterization
of Hedin and Lundqvist.\cite{hedin71} For the bulk geometry we also
report our results obtained with the PBE variant of the generalized
gradient approximation (GGA).\cite{perdew96} The electron-ion interaction was
represented by norm-conserving pseudopotentials in the form proposed by
Vanderbilt,\cite{vande85} treating 6$s$, 6$p$, and 6$d$ as valence states.
The Kohn-Sham orbitals were expanded in a mixed basis (MB) of local
functions and plane waves.\cite{louie79,meyer} Spin-orbit
coupling was incorporated within the pseudopotential scheme via
Kleinman's formulation,\cite{kleinman80} and used to obtain full
charge-self-consistency.\cite{heid10}

For convenience, we have chosen the rhombohedral representation of
the structure of bulk Bi (which contains two atoms per unit cell) and
the hexagonal one \cite{hofmann06} to describe the Bi(111) surface.
The Bi(111) surface was modeled with slabs having one atom per layer
(1$\times$1 in-plane periodicity) and thicknesses of 6 and 12 bilayers. In
the calculations involving the surface, a vacuum of 12--14 \AA\ separates
the periodic images of the slab to avoid interaction between them. We
have used the relaxed system to perform the lattice dynamics calculations.

Integrations over the bulk and surface Brillouin zones (BZ) were
performed by sampling 12$\times$12$\times$12 and 12$\times$12$\times$1
meshes corresponding to 189 and 19 irreducible $k$ points for the bulk
and surface calculations, respectively, combined with a Gaussian broadening
with a smearing parameter of 0.1 eV. The Kohn-Sham orbitals were expanded
in a basis set consisting of local $s$ and $p$-type functions augmented
by plane-waves with a kinetic energy cut-off of 12 Ry (163 eV). The
Fourier expansion of the crystal potential and charge density has been
found to be safely truncated at 50 Ry. The positions of all atoms in the
slab were optimized until the forces on each atom and each direction
was smaller than $10^{-3}$ Ry (2$\times$10$^{-2}$ eV/\AA). For this
purpose, the Broyden-Fletcher-Goldfarb-Shanno algorithm \cite{press92}
has been applied.  
The choice of the parameters has been verified through preliminary
tests on the lattice parameters, band structure, volume, bulk modulus,
and phonon frequencies at some high-symmetry points of bulk Bi. These
tests also showed that the LDA is more appropriate than GGA.

The calculation of the lattice dynamical matrices at specific $q$
points of the surface BZ (SBZ) was performed using linear response theory
embodied within DFPT. \cite{zein84,baroni87} Its implementation in the MB
scheme is described in Ref.~\onlinecite{heid99}. The dynamical matrices
for bulk Bi and Bi(111) slabs were calculated at 189 and 7
irreducible $q$ points, respectively. Real-space force constants were
obtained by taking the standard Fourier transform of the corresponding
dynamical matrices.\cite{gianozzi91} The force constants calculated for
the Bi(111) slabs were then combined with those of bulk Bi to model the
dynamics of much thicker slabs.\cite{heid03} The force constants for
these thick slabs were used to derive the eigenvectors and frequencies
of the vibrational eigenmodes of the slab at arbitrary $q$ points of the
SBZ. Finally, electron-phonon coupling matrix elements were calculated
directly from quantities obtained within DFPT including the contribution
from spin-orbit interaction.\cite{heid10}

\section{Results and Discussion}
\label{sec_results}

Before discussing the properties of the Bi(111) surface, we first present
our results for the structural, lattice dynamical and electron-phonon
coupling properties of bulk Bi.

\subsection{Bulk Bi}
\label{subs_bulk}

\begin{figure}
\includegraphics[width=0.8\linewidth,clip=true]{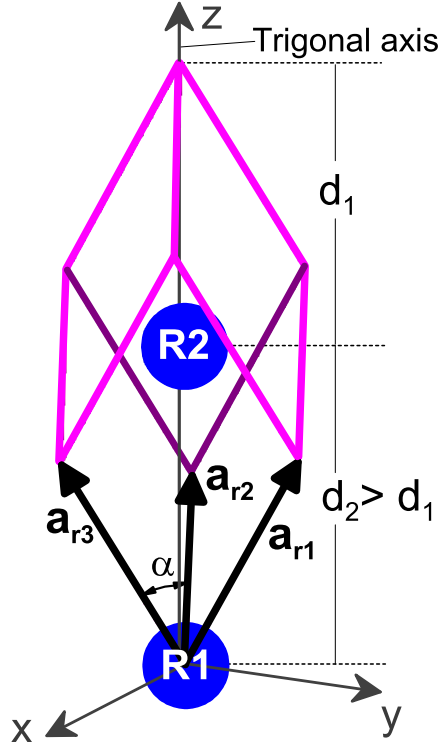}

\caption {(Color online): The structure of bulk Bi in the rhombohedral representation
($a_r$, $\alpha$, and $u$ are given in Table~\ref{tabstruc}), where
${\bf a}_{r1}=b\hat{y}+c_r\hat{z}$,
${\bf a}_{r2}=-b(\sqrt{3}/2\hat{x}+1/2\hat{y})+c_r\hat{z}$,
${\bf a}_{r2}=b(\sqrt{3}/2\hat{x}-1/2\hat{y})+c_r\hat{z}$;
$b=a_r\sqrt{2(1-\cos{\alpha})/3}$;
$c_r=a_r\sqrt{(1+2\cos{\alpha})/3}$;
$d_1=6uc_r$; $d_2=3c_r-d_1$. 
R1 and R2 represent the positions of the two atoms. 
}

\label{figstruc}
\end{figure}

\begin{table}
   \caption{
Structural parameters of bulk Bi. Compared are the present results with
a previous calculation and experimental data (obtained at $T=4.2$\,K).
$d_{\rm short}$ and $d_{\rm long}$ denote the short and long
interlayer distances, respectively.
 \label{tabstruc}
      }
   \begin{ruledtabular}
\begin{tabular}{cccccc}
          & $a_r$ (\AA) & $\alpha$ & $u$ & $d_{\rm short}$ (\AA) & $d_{\rm long}$ (\AA) \\
\hline
LDA &  4.67 & 57.94 & 0.236 & 1.606 & 2.267 \\
GGA &  4.93 & 56.18 & 0.232 & 1.621 & 2.513 \\
LDA [\onlinecite{diaz07}]  &  4.69 & 57.57 & 0.234  & 1.576 & 2.326 \\
\hline
Exp. [\onlinecite{schiferl68}]   &  4.7236(5)& 57.35(1) & 0.23407(4) & & \\
\end{tabular}
   \end{ruledtabular}
\end{table}

The A7 structure that characterizes Bi is a rhombohedral structure
containing two atoms per primitive cell. In describing the crystal, one of
them (R1) can be considered to lie at the origin and the other one (R2)
along the trigonal axis (z-axis), as shown in Figure~\ref{figstruc}. The
rhombohedral representation of the lattice is completely determined by
three parameters, the length of its basis vectors, $a_r$, and the two
parameters gauging its departure from the simple cubic structure, \cite{dresselhaus71}
$\alpha$ and $u$.  $\alpha$ denotes the angle between two
rhombohedral basis vectors, while $u$ is the internal parameter defining
the position of R2 in the primitive unit cell.  The crystal can also be
viewed as being built from hexagonal planes stacked along the trigonal
axis with two alternating distances, $d_{\rm short}$ and $d_{\rm
long}$. In terms of the rhombohedral parameters they are given
by $d_{\rm short}=(6u-1)c_r$ and $d_{\rm long}=(2-6u)c_r$, respectively, where
$c_r=a_r\sqrt{(1+2\cos{\alpha})/3}$ is the length of
the basis vectors projected onto the trigonal axis (see Fig.~\ref{figstruc}).
Their ratio is solely determined 
by $u$ via $d_{\rm short}/d_{\rm long} = (6u-1)/(2 - 6u)$.  Our
results for the lattice optimization are displayed in Table~\ref{tabstruc}
together with those from measurement and previous calculations.
\begin{figure}
\includegraphics[width=0.6\linewidth,clip=true]{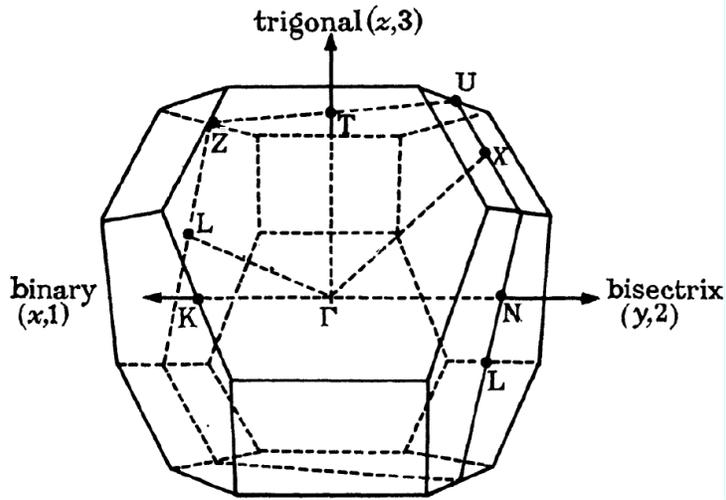}

\caption{The Brillouin zone of the A7 structure of bulk Bi}

\label{figbz}
\end{figure}

\begin{figure}
\includegraphics[width=0.75\linewidth,clip=true]{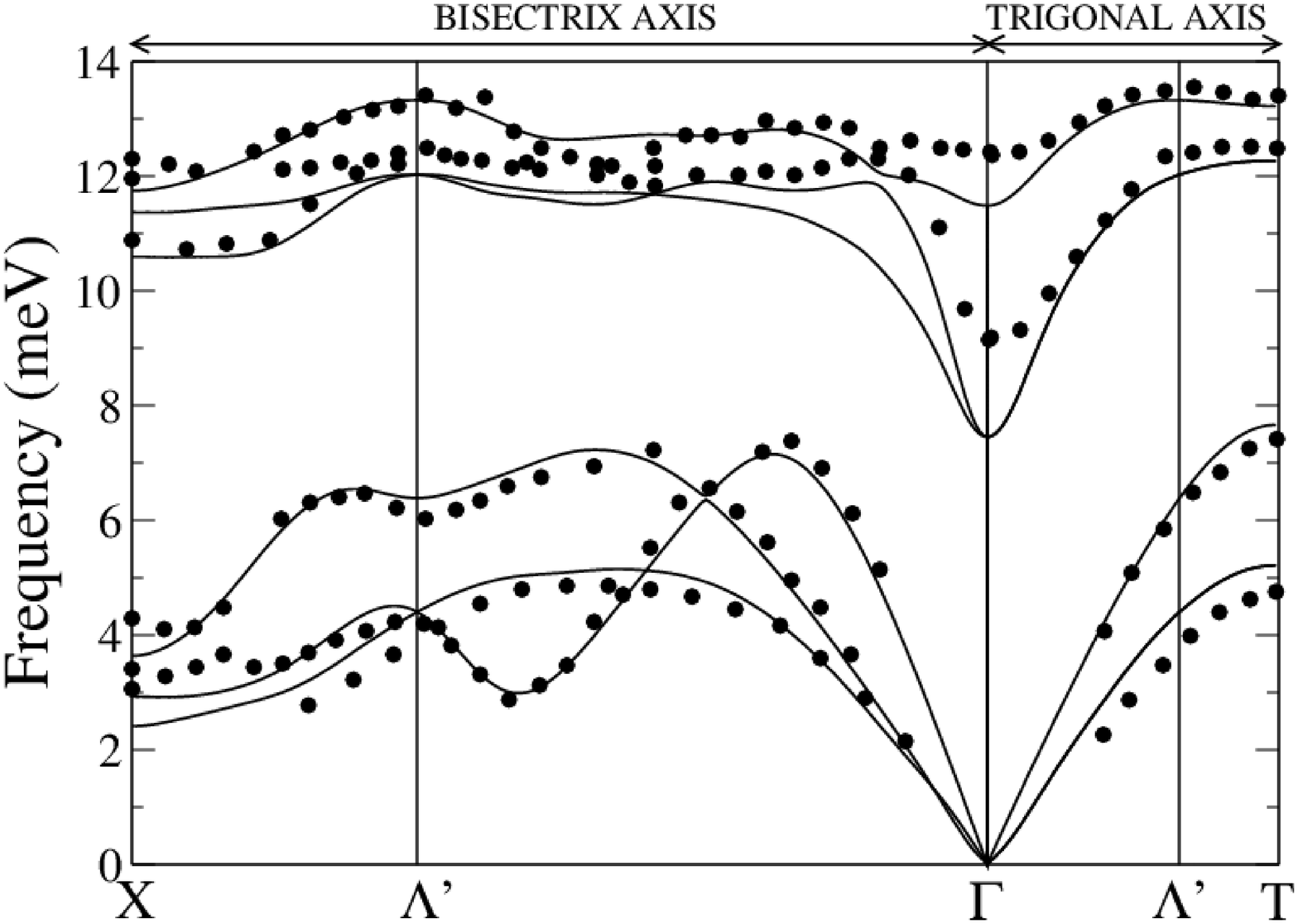}

\includegraphics[width=0.75\linewidth,clip=true]{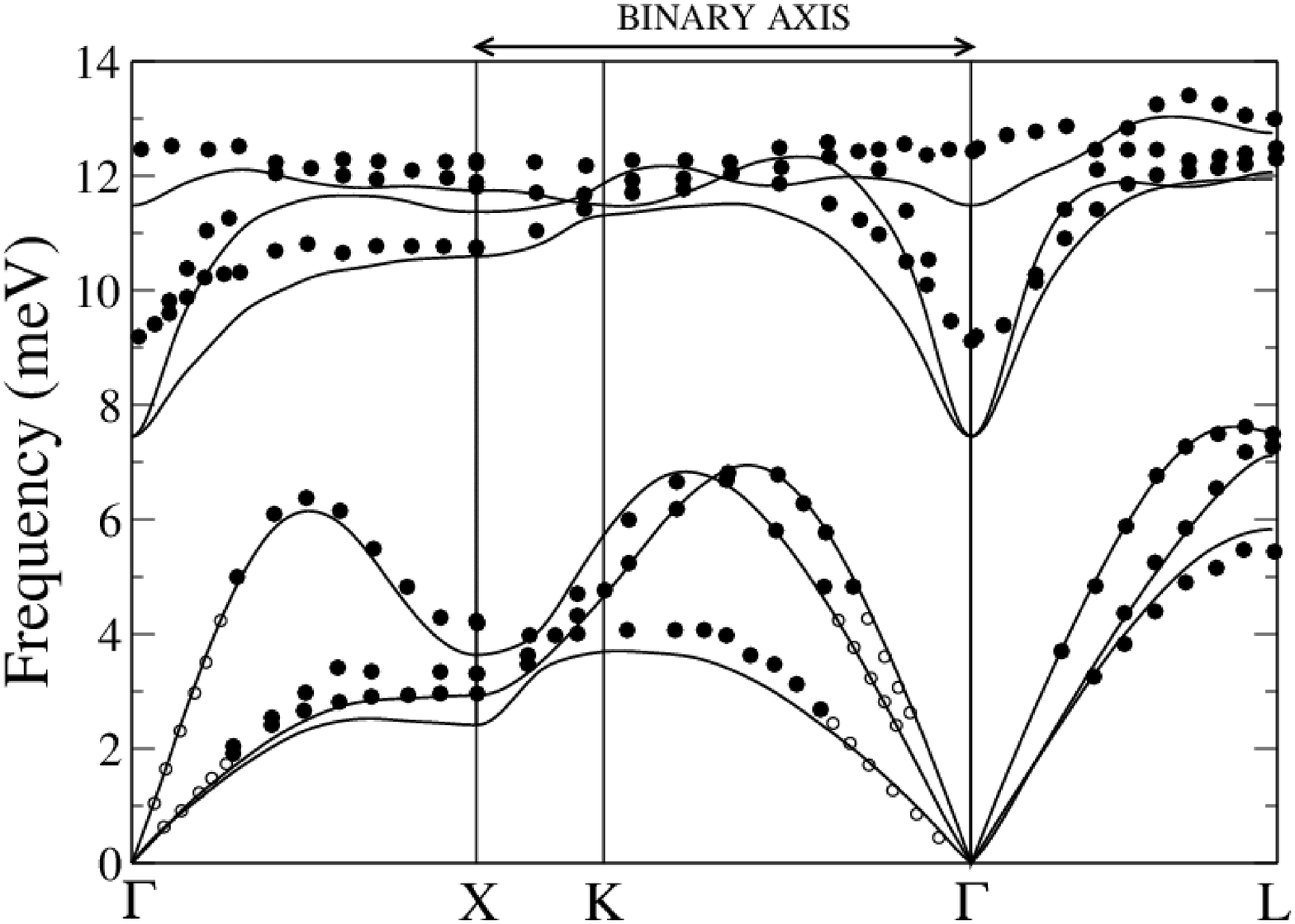}
 
\caption{
Calculated phonon dispersion curves of bulk Bi along high-symmetry
directions of the Brillouin zone. The black dots correspond
to inelastic neutron scattering data taken at 75\,K.\cite{farlane71}
The point $\Lambda^\prime$ lies on the trigonal axis with
${\bf k}_{\Lambda^\prime}=2/3 {\bf k}_{\rm T}$.
}

\label{figphobulk}
\end{figure}

The A7 BZ is a distorted truncated octahedron (see Fig.~\ref{figbz}). The
distortions are relative to the trigonal direction ($\Gamma$T).  Six of
the eight hexagonal facets are irregular and the squared facets are
rather isosceles trapezoids. The regular hexagons contain the T-points
and the irregular ones the L-points. 
It is customary to consider a binary-bisectrix-trigonal system as
indicated in Fig.~\ref{figbz}.  The high-symmetry directions $\Gamma$T ,
$\Gamma$L, $\Gamma$X, and the bisectrix direction lie in the mirror plane
of the A7 structure, while the binary direction is perpendicular to it. 

Figure~\ref{figphobulk} shows our calculated phonon dispersion curves
along the bisectrix, trigonal, $\Gamma$X, binary, and $\Gamma$L
directions together with the most comprehensive experimental data
set by MacFarlane \cite{farlane71} for comparison.  The agreement with
experimental data is quite good at most points in the BZ, except for the
frequencies of the optical modes around the zone center. This also leads
to some discrepancies for the lower optic branches throughout the zone,
while the acoustic branches are well reproduced.

These findings are in line with the previous theoretical study of the
lattice dynamics of Bi by D\'{i}az-S\'{a}nchez {\em et al.}\cite{diaz07}
Comparing calculations without and with SOC, they found that SOC clearly
improves the description of the phonon dispersion, but with the exception
of the optical branches near the $\Gamma$ point.  Overall the performance
of DFT is poorer when compared with calculations for other heavy metals
that are also largely influenced by the SOC.\cite{heid10}

A prominent feature of both experimental and theoretical spectra
is the pronounced dip in the dispersion of the two lower optical
branches at $\Gamma$, with a minimum frequency of about 9~meV
found by inelastic neutron scattering and Raman experiments at low
temperatures.\cite{farlane71,hoehne77} 
Such dips also occur in other group V elements (As, Sb) as
well as in IV-VI compounds like SnTe, and have been interpreted as
caused by long-ranged interactions due to a resonant bonding effect in
materials with pseudo-rocksalt structure.\cite{lee14}
Our calculation
exaggerates this dip, a phenomenon well known from previous theoretical
work.\cite{diaz07} We have checked that including the 5$d$ semicore
states in the valence space does not alter this feature.  Furthermore,
it was not changed by
varying the Gaussian broadening, which indicates that Fermi surface
related renormalization is not responsible for it.  
We observed, however, a high sensitivity on
changes of the lattice structure, in particular on the internal structural
parameter $u$. Using the experimental value of $u=0.234$ instead of the
optimized value $0.236$ raises the lower and upper optical frequencies
at $\Gamma$ by 1.2\,meV and 0.8\,meV, respectively, thereby reducing the
difference to the measured frequencies to less than 0.6\,meV. Similar
improvements are found for all optic branches.

We have performed calculations of the electron-phonon coupling
constant in bulk Bi, given by
\begin{equation}
  \lambda=\frac{1}{N(E_F)} \sum_{{\bf q}\nu}
\sum_{{\bf k}ij}
\frac{2}{\omega_{{\bf q}\nu}}
|g^{{\bf q}\nu}({{\bf k+q}j, {\bf k}i)}
 |^2 \delta(\epsilon_{{\bf k}i}-E_F) \delta(\epsilon_{{\bf
 k+q}j}-E_F)
 \,.
\label{eqlambda}
\end{equation}
Here, $g$ denotes the screened electron-phonon coupling matrix elements,
and the sum runs over all phonon modes (with momentum ${\bf q}$,
branch index $\nu$, and frequency $\omega_{{\bf q}\nu}$) and over all
electronic states (with momentum ${\bf k}$, band index $i$, and energy
$\epsilon_{{\bf k}i}$). $N(E_F)$ denotes the electronic density of states
per spin at the Fermi energy $E_F$. Since the bulk Fermi surface consists
of tiny Fermi pockets, care must be taken in the BZ sampling to assure a
proper convergence. We have used meshes of up to 36$\times$36$\times$36
$k$ points for the electronic states and approximated the delta functions
by Gaussians with widths of 0.1--0.2 eV. For the sum over phonon modes,
the 12$\times$12$\times$12 $q$-point mesh was applied. In all cases,
we found $\lambda<0.1$, which is consistent with the absence of
superconductivity in bulk Bi down to 50 mK.\cite{tian08} In a recent
linear-response calculation using GGA a very similar value of 0.09 was
found.\cite{huang13}

\subsection{Bi(111) geometry}
\label{subs_111geo}

\begin{figure}
\includegraphics[width=0.90\linewidth,clip=true]{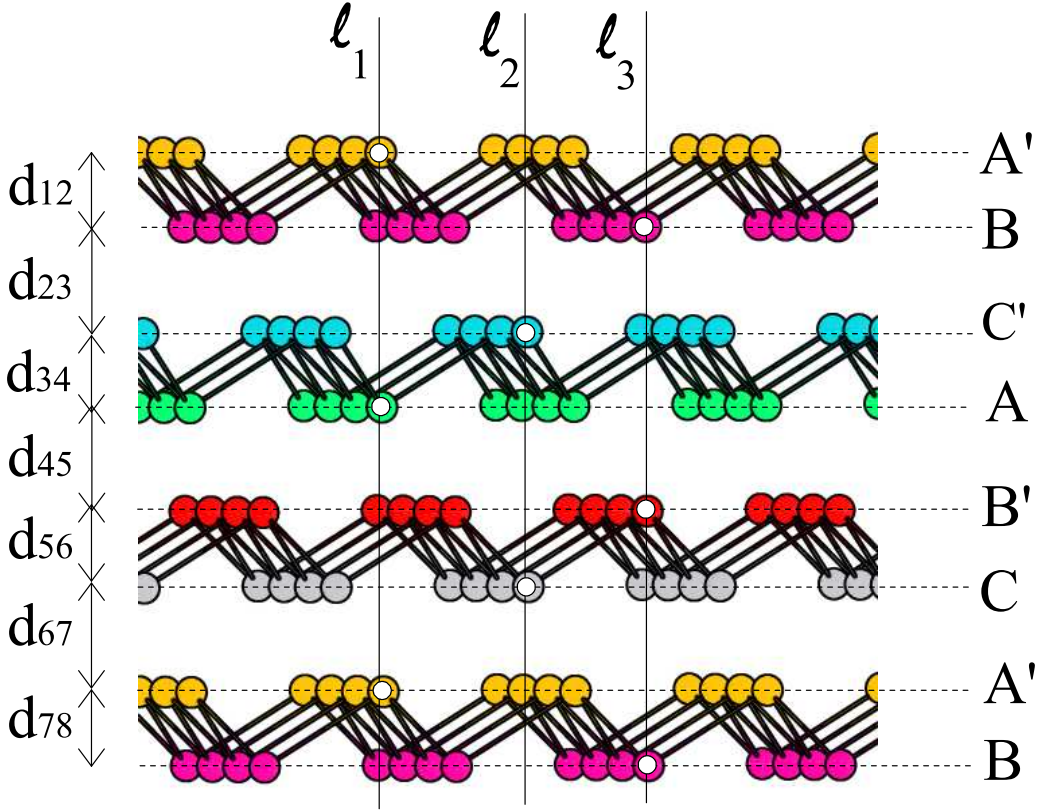}

\caption{(Color Online) Side view of a sketch of
the Bi(111) structure. A, A$^\prime$, B, B$^\prime$, C, and C$^\prime$
represent the planes defined by each layer. The sticks connecting the
balls are used to indicate only first nearest neighbors (between atoms on planes
A$^\prime$ and B, C$^\prime$ and A, or B$^\prime$ and C) to highlight
the bilayer-like structure of Bi(111). 
$d_{ij}$ denotes the vertical spacing between adjacent planes.
}
\label{fig111struc}
\end{figure}

The (111) surface of Bi is that perpendicular to the trigonal axis.
The surface can, in principle, be formed by breaking either the three
bonds of each atom with its 2nd nearest neighbors (NN) or those with its 1st NNs. The latter
case, however, would require much more energy than the former. 
Its low reactivity\cite{jona67} also speaks in favor of breaking
the weaker bonds with 2nd NNs.
As such, Bi(111) can be considered as formed by
a set of bilayers (see Fig.~\ref{fig111struc}) where intra-bilayer bonds
(between 1st NNs) are rather covalent (strong) and the vertical intra-bilayer spacing is
relatively short ($d_{12}$, $d_{34}$, etc.) whereas inter-bilayer bonds
(between 2nd NNs) are rather metallic (weak) and the vertical inter-bilayer spacing
is relatively long ($d_{23}$, $d_{45}$, etc.). 
They correspond to the interlayer distances $d_{\rm short}$ and 
$d_{\rm long}$ of the bulk, respectively.
Figure~\ref{fig111struc}
shows that the stacking of planes in an ideally bulk-terminated Bi(111)
surface (before relaxation) is such that every six planes the structure
repeats itself. Each atom in any layer has three 1st NNs and three 2nd
NNs in the two adjacent layers.  There are three kinds of layers A(A'),
B(B'), and C(C'). A and A' layers have atoms lying on the $l_1$ line,
B and B' layers have atoms lying on $l_2$, and C and C' have atoms
lying on $l_3$.  The stacking makes Bi(111) a quite open surface. The
closest distance between atoms within a layer is ~4.54 \AA\, such that
the surface leaves exposed even the third layer.

\begin{table}
   \caption{
Changes (\%) in the interlayer spacing between the layers of clean
Bi(111) films with respect to that in bulk Bi, $\Delta d_{i(i+1)}$.
Given are the present results for three slabs of 6, 10, and 12 bilayers (BL).
Changes in deeper interlayer spacings are below 0.1\%. Shown are
also values obtained in previous theoretical works and an
experimental study. T$\to$ 0 K values are estimates based on
extrapolated values obtained at 140 K and higher temperatures.
 \label{tab111struc}
      }
   \begin{ruledtabular}
\begin{tabular}{cccccccc}
&\multicolumn{5}{c}{Theory}&\multicolumn{2}{c}{Experiment\footnote{low-energy 
electron diffraction, Ref.~\onlinecite{moenig05}}}\\
 & 6~BL & 10~BL & 12~BL & 
6~BL\footnote{GGA, pseudopotential plane-wave method, Ref.~\onlinecite{tamtoegl13}} & 
7~BL\footnote{LDA, full-potential linearized augmented plane wave method, 
Ref.~\onlinecite{moenig05}} & T=140 K & T$\to$0 K \\ 
\hline
 $\Delta d_{12}$ & -1.62 & -1.56 & -1.56 & -0.83 & +0.6 & +0.5$\pm$1.1 & +1.2$\pm$2.3 \\
 $\Delta d_{23}$ &  1.37 &  1.32 &  1.32 & +3.13 & +6.2 & +1.9$\pm$0.8 & +2.6$\pm$1.7 \\
 $\Delta d_{34}$ & -0.87 & -0.93 & -0.87 &       &      &  0.0$\pm$1.1 &  \\
 $\Delta d_{45}$ &  0.13 &  0.18 &  0.13 &       &      &              &  \\
 $\Delta d_{56}$ & -0.25 & -0.31 & -0.31 &       &      &              &  \\
\end{tabular}
   \end{ruledtabular}
\end{table}

In bulk Bi, the intra-bilayer distance is given by $d_{\rm short}=1.606$~\AA, while the
vertical inter-bilayer distance is $d_{\rm long}=2.267$~\AA.  Upon the creation of the
surface the interlayer distances relax. Table~\ref{tab111struc} shows the calculated
relaxation for three different thicknesses of the slab (6, 10 and 12
bilayers) together with previous calculations\cite{tamtoegl13,moenig05}
and low-energy electron diffraction (LEED) measurements.\cite{moenig05}
Our results show a fast convergence of the relaxation with the slab
thickness. Already for a 10-bilayer slab the interlayer distances are
sufficiently converged, and the bulk spacing at the center of the slab
is recovered.  We found a shrinking of the first intra-bilayer distance
($d_{12}$) of 1.56\% and an expansion of the first inter-bilayer distance
($d_{23}$) of 1.32\%.  The first result is at variance with the LEED
experiment, where an expansion of $d_{12}$ was found, although the
experimental error bars are large.  Our value for $d_{23}$, however,
agrees with the LEED result.  Our pseudopotential results contrasts a
previous calculation using the full-potential linearized augmented plane
wave method, where both distances were found to expand,\cite{koroteev08}
but a very large expansion of ~6-7\% was predicted for $d_{23}$ at
variance with experiment.  A recent pseudopotential calculation employing
GGA found relaxations much closer to the present ones.\cite{tamtoegl13}

\subsection{Bi(111) electronic structure}
\label{subs_111el}

\begin{figure}
\includegraphics[width=0.4\linewidth,clip=true]{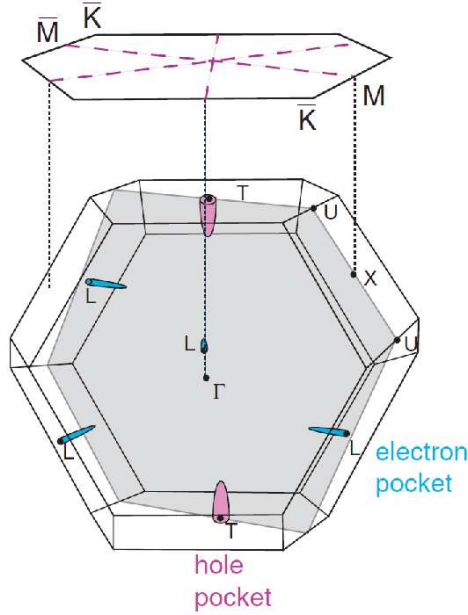}

\caption{(Color Online) The projection of the bulk Brillouin zone of Bi 
onto the (111) surface. Sketched are the bulk Fermi surfaces
(exaggerated in size). Figure is taken from Ref.\,\onlinecite{hofmann06}.
The shaded area denotes the mirror plane.  }

\label{figsbz}
\end{figure}
\begin{figure}
\includegraphics[width=0.9\linewidth,clip=true]{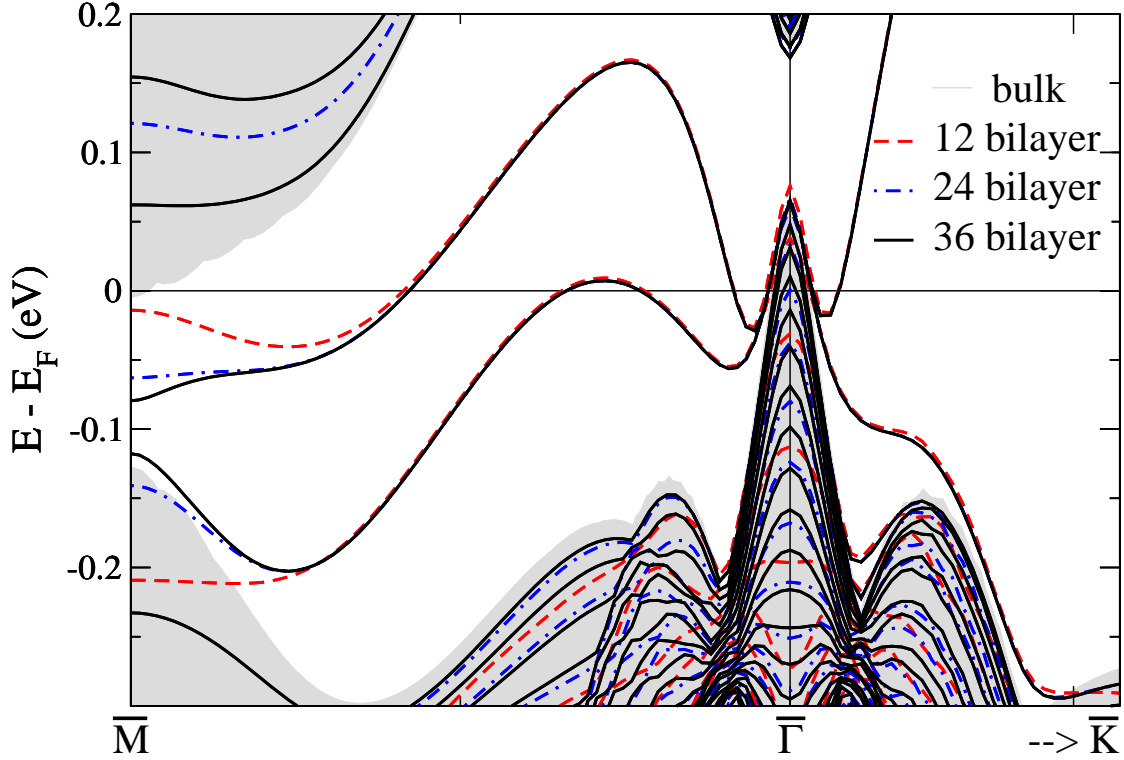}

\caption{(Color Online) Surface bandstructure of a Bi(111) slab.
Shown are the electronic bands in the vicinity of the Fermi energy for
the 12~BL, 24~BL, and 36~BL slabs, respectively.
The grey areas represent regions of bulk states.
}

\label{fig111edisp}
\end{figure}

As indicated in Fig.~\ref{figsbz}, the tiny electron and hole
pockets of bulk Bi are projected onto regions around the $\overline{\rm M}$
and $\overline{\rm \Gamma}$ points of the (111) surface BZ, respectively.
In between, there are larger regions where the bulk spectrum possesses an energy gap near
the Fermi level. As shown in Fig.~\ref{fig111edisp}, the electronic
bandstructure of Bi(111) exhibits two surface localized bands along
$\overline{\rm \Gamma} \overline{\rm M}$ and part of the $\overline{\rm
\Gamma} \overline{\rm K}$ line, which fall into this energy gap region.
They are derived from two semi-relativistic bands, one from each side of the slab.
Since
the surface electronic bands show a rather poor convergence with slab
thickness, we performed calculations of the electronic bandstructure for
12, 24, and 36 bilayers of Bi(111).  As shown in the previous Subsection,
these thicknesses are sufficient to converge the interlayer distances at
the surface and to recover the bulk spacing in the center of the slabs.
In contrast, the surface electronic bands are not converged, but exhibit a
sizable dependence on the film thickness in particular when approaching
the $\overline{\rm M}$ point.  
It was argued that for a semi-infinite surface, the two
split bands should be degenerate at $\overline{\rm M}$ on symmetry
grounds.\cite{hirahara06,koroteev08}
However, in the case of finite slabs such a degeneracy is not
found. There is a strong hybridization
between the upper and lower surface that does not allow convergence of the
energy of these bands around $\overline{\rm M}$ even for slabs of 36
bilayers or more.\cite{koroteev08}
In contrast, this inter-surface
interaction is much weaker away from $\overline{\rm M}$, such that both
electron and hole pockets are well converged already for a 12-bilayer
Bi(111) slab.  
A more detailed analysis of the electronic structure and its
dependence on the slab thickness will be published elsewhere.\cite{alcantara}
Due to such a slow convergence, we
will restrict the analysis of the electron-phonon induced self-energy
effects (Sec.~\ref{subs_epc}) to those parts of the $\overline{\rm M}
\overline{\rm \Gamma} \overline{\rm K}$ lines, where the surface band
energies are sufficiently converged for a 12-bilayer slab.

\subsection{Bi(111) surface phonons}
\label{subs_111pho}

\begin{table}
   \caption{
Surface localized phonons of Bi(111) at high symmetry points of the
surface Brillouin zone for a slab of 50 BL obtained with the slab
filling procedure.  Frequencies (in meV) and dominant polarization
are shown with respect to the propagation vector.  For modes at
$\overline{\Gamma}$ and $\overline{\rm M}$, the polarization is
indicated by SV for shear vertical, L for longitudinal and SH for shear
horizontal, while the subscripts indicate the layers with larger weight,
starting with the outermost surface layer.  RW denotes the Rayleigh wave.
For $\overline{\rm K}$, modes show a more complex polarization pattern,
which is therefore not shown.  Our values are compared with available
data from helium atom scattering experiment and theoretical slab
calculations. In the latter cases, no slab filling was performed.
 \label{tab111freq}
      }

\begin{ruledtabular}
\begin{tabular}{ccccc}
Point & This work &  Experiment\footnote{Helium atom scattering data 
from Tamt\"ogl {et al.}, Ref.~\onlinecite{tamtoegl13}, taken from their 
low-temperature data of Fig.~4(a)} 
    & \multicolumn{2}{c}{Theory} \\
&&  &  5 BL \footnote{LDA pseudopotential calculation 
             by Chis {et al.}, Ref.~\onlinecite{chis13}.  Values shown for 
             $\overline{\rm M}$ are estimates based on their Fig.~3}  
    &  6 BL \footnote{GGA pseudopotential calculation without 
             spin-orbit interaction, Ref.~\onlinecite{tamtoegl13}} 
\\ \hline
$\overline{\Gamma}$ 
&12.3 (L$_{1-6}$)              &               & 11.8, 12.2     &  \\
&12.3 (SH$_{1-6}$)             &               &      &  \\
&13.3 (SV$_{3-6}$)             &               &      & 14.2 \\
&13.6 (SV$_{1,2}$)             &  14-15        & 13.6     & 15.3  \\
$\overline{\rm M}$ 
& 2.9 (RW, SV$_{2}$+L$_{1}$) & $\approx 3$ \\
&11.5 (SV$_{1,2}$+L$_{1,2}$) &               & $\approx 11.5$\\
&12.7 (SV$_{3-8}$+L$_{3-8}$) &     \\
&12.9 (SV$_{1,2}$+L$_{2}$)   & $\approx 14$  & $\approx 13$\\
$\overline{\rm K}$ 
& 3.4 (RW)                     & $\approx 3.8$\\
&4.1, 4.6, 4.8, 5.9, 6.4  \\
&11.1, 11.4, 11.8, 12.2      &$\approx 12.3$\\
\end{tabular}
\end{ruledtabular}
\end{table}

\begin{figure}
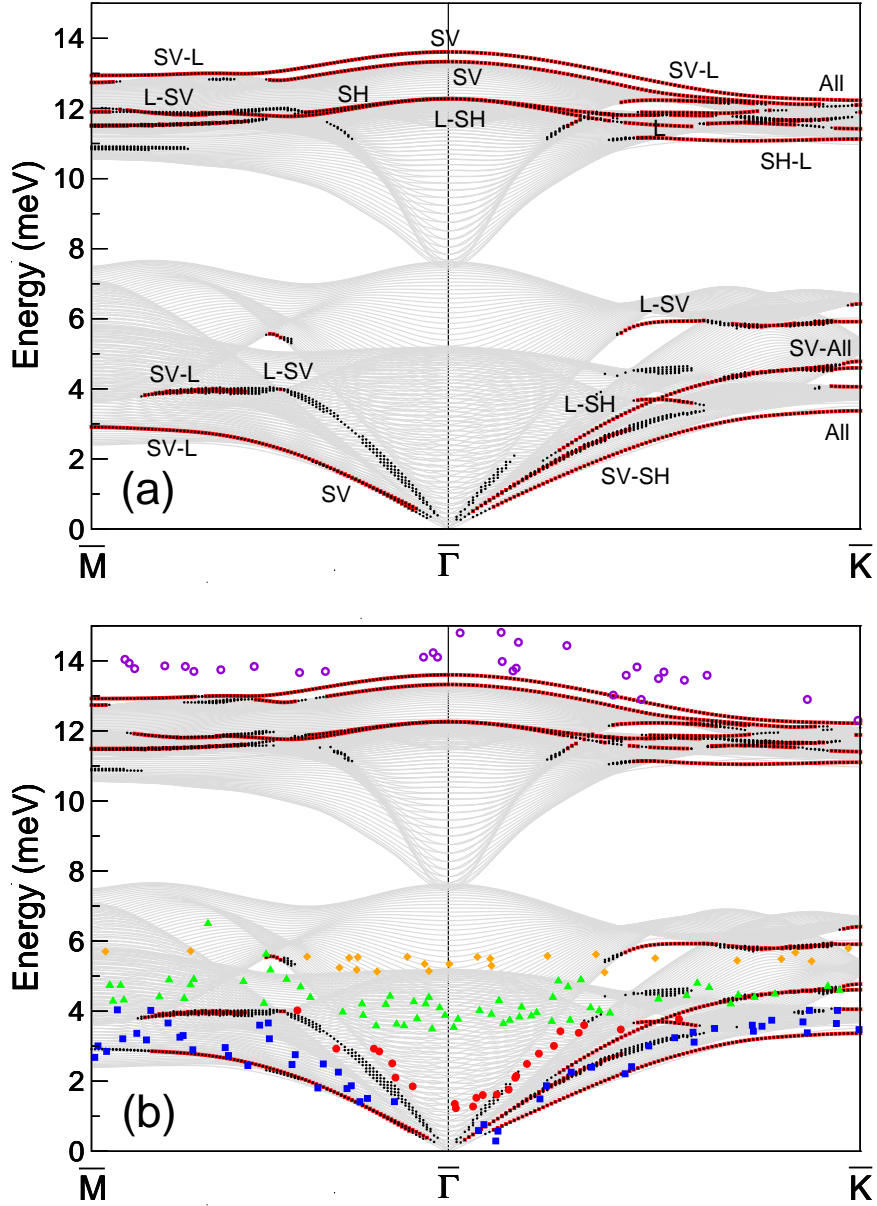

\includegraphics[width=0.70\linewidth,clip=true]{fig7a.eps}

\includegraphics[width=0.70\linewidth,clip=true]{fig7b.eps}

\caption{(Color Online) Phonon dispersion curves of a Bi(111) slab
of 50 bilayers along the high-symmetry directions of the surface
Brillouin zone obtained by slab-filling of (a) a 6-bilayer and (b) a
12-bilayer slab. The dots correspond to surface modes whose amplitude
weight in the two outermost bilayers are at least 20\% (red) or 10\%
(black).  The light grey lines are bulk modes.  In (a) the polarization
of the surface modes away from the high symmetry points is succinctly
indicated with SV for shear vertical, L for longitudinal and SH for shear
horizontal or the word ''All'' if the polarization is not well defined.
In (b) symbols correspond to peaks in inelastic Helium atom scattering
spectra taken at lower temperatures as reported by Tamt\"{o}gl {\em
et al.}\cite{tamtoegl13} }

\label{fig111pdisp}
\end{figure}

We now address the lattice dynamics of the Bi(111) surface.  Recently,
three theoretical studies were devoted to surface phonons in the
framework of the density functional theory.  \cite{chis13,huang13,yang13}
Due to the heavy numerical work involved in dealing with SOC, these
calculations were performed only for thin slabs up to 6 bilayers, where
the vibrational spectrum exhibited significant variations as a function
of thickness. As we are primarily interested in the properties of a
semi-infinity Bi(111) surface we checked the convergence of the surface
localized vibrations by performing lattice dynamics calculations for 6-
and 12-bilayer slabs, respectively. In both cases, the slabs were relaxed
and SOC was taken into account. Surface phonon spectra were then obtained
by combining the real-space force constants of the slab with those from
the bulk calculation to simulate the dynamics of an asymmetric slab of
50 bilayers. An asymmetric slab contains one surface taken from the
fully relaxed slab, while the other surface corresponds to the ideal
bulk-truncated case. The phonon spectrum of such a slab thus contains
surface modes related to both types of surfaces. From the components of
each eigenvector, one can determine the vibrational weight at each atomic
site and at each direction.  Fig.~\ref{fig111pdisp} shows the phonon
dispersions based on the 6 and 12-bilayer calculations, respectively.
The optic and acoustic parts of the projected bulk bands (grey lines)
slightly touch each other at $\overline{\Gamma}$, because the lower
optic bulk mode at the zone center has such a small calculated frequency.
Surface localized modes of the true (relaxed) surface are identified
by their vibrational weight in the first two bilayers at the surface.
Modes with a weight of more the 20\% are highlighted in red, while
those with a weight of more the 10\% are indicated by black dots.  A more
detailed characterization of surface modes at the high-symmetry points
$\overline{\Gamma}$, $\overline{\rm M}$, and $\overline{\rm K}$ is given
in Table~\ref{tab111freq}.

One can note that after the slab filling procedure, both calculations
give essentially the same surface localized spectrum. This indicates
that changes in the real-space force constants due to the presence of
the surface are essentially confined to the outer two or three bilayers
and are reasonably well converged already for the 6-bilayer slab. The
fast convergence of the dynamical properties contrasts the very slow
convergence of the surface electronic band structure. This indicates that
the uncertainty in the surface band structure near the $\overline{\rm
M}$ point has little influence on the electronic response to atomic
perturbations and hardly affects the frequencies of the surface phonons.

A prominent feature of the Bi(111) surface phonon dispersion is the
appearance of modes above the optic bulk band almost everywhere along the
$\overline{\rm KM}$ and $\overline{\Gamma \rm M}$ directions.  The upper
branch consists of predominately vertical vibrations of atoms in the
first bilayer, and reflects a 12\% stiffening of the force constants
between the two layers forming the first bilayer and the contraction
of the corresponding interlayer spacing (see Table~\ref{tab111struc}).
The second highest branch above the optic bulk spectrum also has a shear
vertical polarization but is mainly localized in the 2nd and 3rd bilayer.
A third surface branch with in-plane (shear-horizontal and longitudinal)
polarization is present within the optic bulk spectrum and is localized
over the first three bilayers.

In the acoustic spectrum, the Rayleigh wave (RW) falls slightly below the
bulk band along $\overline{\Gamma \rm K}$, while it remains within the
bulk continuum along $\overline{\Gamma \rm M}$.  Along $\overline{\Gamma
\rm K}$, above the RW, there appear several modes localized primarily
in the first bilayer.  The surface mode at about 6\,meV stretching from
$\overline{\rm K}$ halfway to $\overline{\Gamma}$ involves vibrations of
atoms in the 2nd bilayer with mainly shear vertical and to a lesser extend
in-plane polarization.  Along the $\overline{\Gamma \rm M}$ directions,
the longitudinal 1st-bilayer mode lying above the RW shows a weaker
localization near $\overline{\Gamma}$, but when approaching $\overline{\rm
M}$ it becomes more localized and acquires a stronger vertical component.

The surface phonons have been recently investigated via inelastic helium
atom scattering (HAS) by Tamt\"{o}gl {\em et al.}\cite{tamtoegl13} In
Fig.~\ref{fig111pdisp}(b) we have added the reported frequencies taken
at lower temperatures ($T=103/123$\,K).  They clearly observed surface
modes above the optic bulk band and interpreted their data as evidence
for two branches with shear vertical polarization, in agreement with
our calculation.  Their frequencies are about 1\,meV higher than our
theoretical result, but also lies significantly above the experimental
bulk maximum of $\approx$13.5\,meV (see Fig.~\ref{figphobulk}).  In the
acoustic part, good agreement between our calculations and the HAS
results are found for the RW and the longitudinal 1st-bilayer mode as
well  as the 2nd-bilayer mode at about 6\,meV.

In the HAS spectra, a variety of additional peeks were observed in the
range of 4--5\,meV, most prominent near $\overline{\Gamma}$.  They were
interpreted as caused by longitudinally polarized vibrations in the 3rd
layer (2nd bilayer).  This interpretation was based on the theoretical
vibrational spectrum of a 6-bilayer slab and supported by calculations
of the charge density variations induced by such vibrations, which were
found to be large enough at the scattering point of the helium atom to
explain the observed HAS peaks.

When we consider the 12-bilayer slab, we do find similar vibrational modes
in this energy range with enhanced longitudinal polarization in the 3rd
and 4th layer.  However, this enhanced vibrational weight is a property
of thin slabs only.  When simulating the dynamics of thicker slabs
by the slab-filling procedure, the weight becomes successively smaller
with increasing thickness. It remains an open question, if the density
oscillations related to these modes remain strong enough in the limit
of thick slabs to explain the HAS peaks, which have been observed for
Bi(111) samples of macroscopic thickness.

\begin{figure}
\includegraphics[width=0.75\linewidth,clip=true]{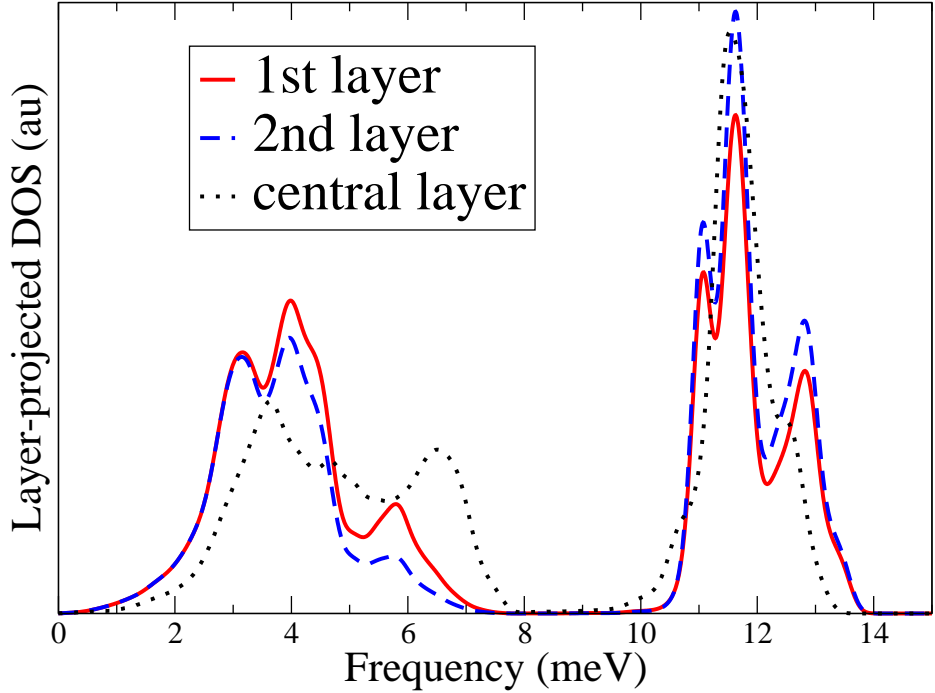}

\caption{(Color Online) Layer-projected density of states (LDOS)
obtained for a 12-bilayer Bi(111) slab. The red (full) and blue (dashed)
lines show the LDOS of the first and second layer of the outermost
bilayer, respectively. The black (dotted) curve corresponds to a central
bilayer and represents a bulk-like spectrum.  }

\label{fig111pdos}
\end{figure}

The presence of low-frequency optical modes causes the surface atoms to
have an enhanced vibrational density of states at lower frequencies.
Fig.~\ref{fig111pdos} shows the layer-projected phonon density of
states (LDOS) for the first, second, and twelfth layer
of the 12-bilayer slab. The twelfth layer is deep enough so that it
undergoes no relaxation (see Table~\ref{tab111struc}) and thus its LDOS
is representative of the phonon DOS of bulk Bi. Fig.~\ref{fig111pdos}
shows that the LDOS of the first two layers have an additional peak at
the upper end of the optical spectrum as compared to the bulk due to
the presence of the super-bulk surface modes. In contrast, the acoustic
part of their LDOS is softened with respect to the bulk, which can be
traced back to a softening of the nearest-neighbor coupling within the
first layer.  For each LDOS curve the Debye-like behavior applies in
the beginning of the spectrum. From the initial quadratic behavior
we deduce a Debye cutoff frequency of 8.9 meV for the bulk
and of 7.3 meV for the first and second surface layer, respectively.
These values are faily consistent with experimental Debye frequencies of
about 10\,meV for the bulk\cite{ast02,ramanathan55} and 6.5\,meV for the
Bi(111) surface layer. The latter value was deduced from measurements
of the mean-square atomic vibrational amplitudes of the surface
atoms.\cite{moenig05} The Debye model is often used as an approximation
for the Eliashberg function describing the coupling of an electronic state
to the phonon system.  We will show in the following Section, however,
that the spectral shape of the coupling function can deviate significantly
from the simple Debye form for the surface electronic states of Bi(111).


\subsection{Electron-phonon interaction}
\label{subs_epc}

The metallic character of the Bi(111) surface gives the opportunity of
studying the $e$-ph interaction  
for states which are well confined at the surface and
subject to a strong spin-orbit interaction.
To quantify the strength of the electron-phonon interaction, the
dimensionless $e$-ph coupling parameter is used: \begin{equation}
\lambda_{\mathbf{k}i} = \int_{0} ^{\omega_{\rm{max}}} \frac{\alpha
^{2} F^{\rm E}_{\mathbf{k}i}(\omega)+\alpha ^{2} F^{\rm
A}_{\mathbf{k}i}(\omega)}{\omega} \mathrm{d}\omega    \ .
\label{eq:lambda}
\end{equation}
Here $\mathbf{k}i$ denotes an electron (hole) state momentum and band
index, $\omega_{\rm{max}}$ is the maximum phonon frequency and $\alpha
^{2} F^{\rm {E(A)}}_{\mathbf{k}i}(\omega)$ is the electronic state
dependent Eliashberg spectral function corresponding to phonon emission
(E) and absorption (A) processes:\cite{grimvall81}
\begin{eqnarray}
\alpha^{2}F^{\mathrm{E(A)}}_{\mathbf{k}i}(\omega) = \sum
_{\mathbf{q}\nu,f} \delta (\epsilon_{\mathbf{k}i}
- \epsilon_{\mathbf{k+q}f}\mp \omega _{\mathbf{q}\nu} ) \nonumber \\
\times \mid \mathrm{g}^{\mathbf{q}\nu}(\mathbf{k+q}f,\mathbf{k}i )|
^{2}\delta (\omega - \omega _{\mathbf{q}\nu} ) \ . \label{eq:alfa}
\end{eqnarray}
The $"-"$ and $"+"$ signs in the delta function with electron energies
correspond to phonon emission and absorption, respectively. The
sum is carried out over final electronic states ($f$) and phonon modes
($\mathbf{q}\nu$).

In the calculation of the $e$-ph coupling a 12-bilayer Bi(111)
film is used, and only those states in the surface electronic bands that
do not differ in energy from those of the 24- and 36-bilayer calculation
are considered, i.e. we restrict our analysis to momenta
in the range $\sim0.6|\overline{\Gamma \rm M}|$).
To see how the surface electronic states couple to phonons we have
calculated $\lambda_{\mathbf{k}i}$ for different electron (hole)
energies and momenta. The summation over phonons in Eq.  \ref{eq:alfa}
was carried out over 1296 wave vectors ($36\times36$) in the surface
BZ. The delta function with electronic energies was approximated by
a first-order Hermite--Gaussian function with a smearing width in the
range of 0.07--0.4 eV.

\begin{figure}
\includegraphics*[width=0.85\columnwidth]{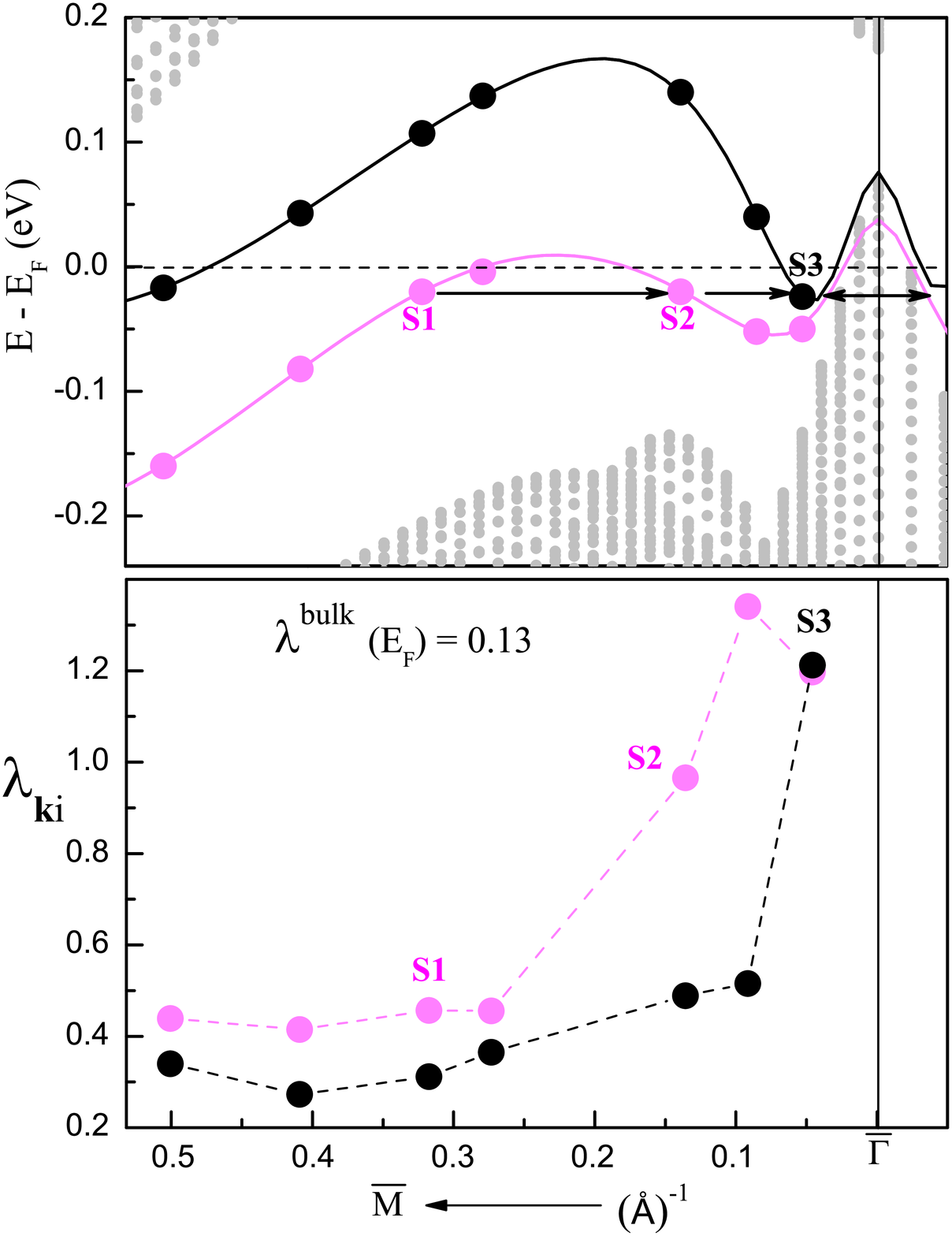}
\caption{(Color Online) (Top) Thick circles indicate the surface
electronic states for which the electron-phonon coupling calculations
were done. S1, S2, and S3 denote
specific surface states of similar binding energies, which are
discussed in the text in more detail. Bulk Bi bands projected onto the
(111) plane are indicated by small gray circles. (Bottom) Electron-phonon
coupling parameter $\lambda_{\mathbf{k}i}$ as a function of electron (hole)
momentum along the $\overline{\Gamma \rm M}$ symmetry direction for
electronic states in the lower (gray/purple) and higher (black)
surface bands. } \label{fig:lambda}
\end{figure}

The electron-phonon parameter $\lambda_{\mathbf{k}i}$ as a function
of momentum is shown in Fig. \ref{fig:lambda} for the two surface
bands.  The electron-phonon
coupling near the Fermi level is of intermediate strength
($\lambda \approx $0.45, the lower state) unless the electronic state
lies close to $\overline{\Gamma}$. It is markedly different from
the very weak coupling constant of $\lambda<0.1$ found at the Fermi
level of bulk Bi (see Sec.~\ref{subs_bulk}).  The strength of $e$-ph
coupling for excited electrons on Bi(111) is also higher
than the values of $\lambda_{\mathbf{k}i}=0.2-0.3$ obtained
for the surface electronic states near the Fermi level on the Bi(100)
and Bi(110) surfaces.\cite{gayone03,kirkegaard05}

\begin{figure}
\includegraphics*[width=0.9\columnwidth]{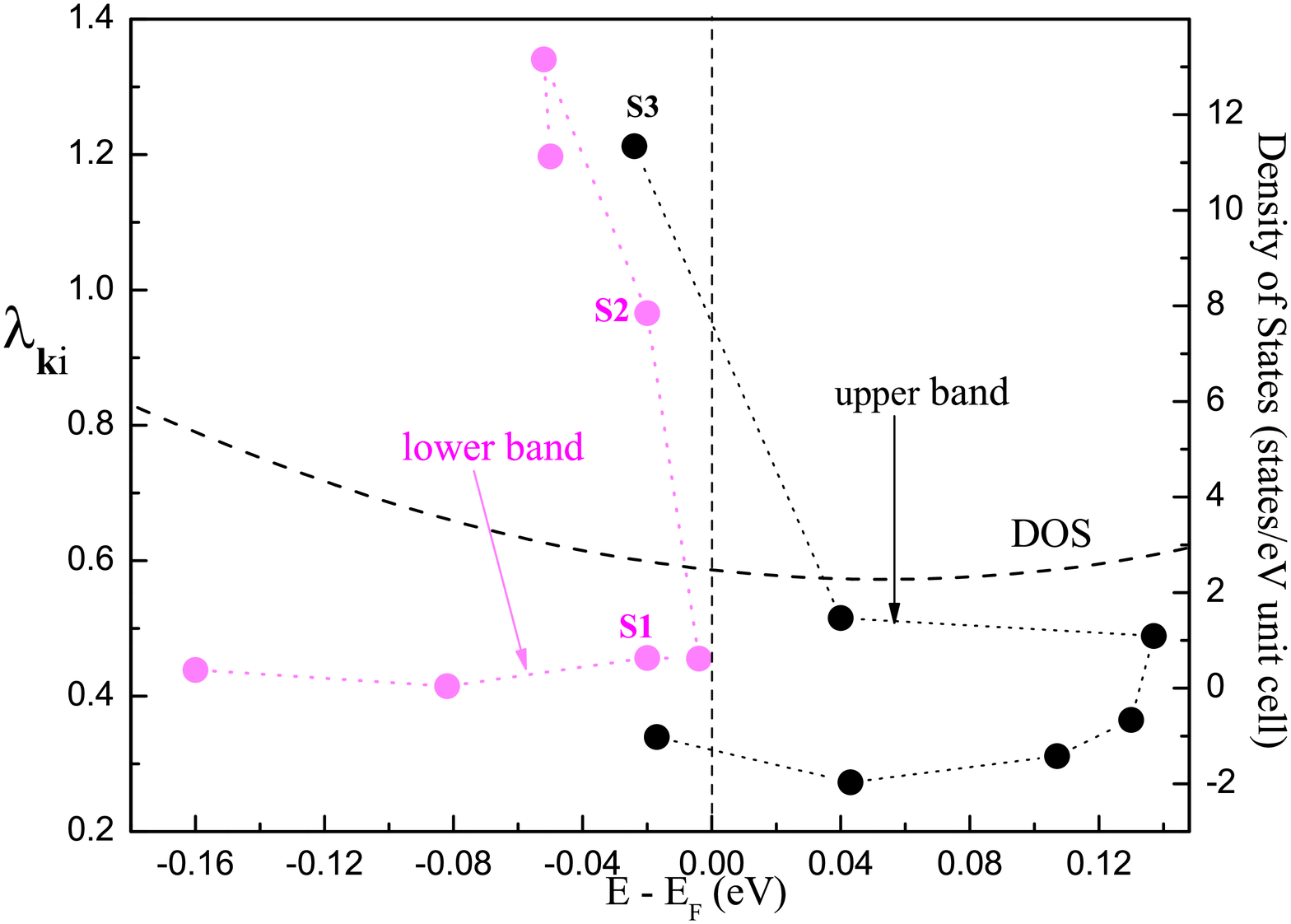}
\caption{(Color Online) Electron-phonon coupling parameter $\lambda_{\mathbf{k}i}$
as a function of electron energy. The notation is the same as in
Fig. \ref{fig:lambda}. The points are connected according to their
positions in the $\overline{\Gamma \rm M}$ direction. The dashed
line shows the calculated density of electronic states, $N(E)$, as a
function of energy.} \label{fig:dos}
\end{figure}

Due to a small energy range covered by the surface states ($\sim$0.15
eV around $E_{\rm F}$) the coupling strength is not expected to vary
considerably with electron momentum if the scattering within the surface
bands dominates.  That is the case at large electron momentum. As is
evident from Fig.~\ref{fig:lambda}, for $|\mathbf{k}|>0.2|\overline{\Gamma
\rm M}|$ the calculated $\lambda_{\mathbf{k}i}$ shows a weak dependence
on electron momentum and the strength of the $e$-ph interaction varies
between 0.3 and 0.5 for both electronic bands, i.e. it is not large
for either of the two. There is also an indication of a rather weak
dependence on the electron binding energy, namely, the difference in
the coupling strength between the two electronic bands at the same
$\mathbf{k}$ is small and does not exceed 0.15. In order to illustrate
this $\lambda_{\mathbf{k}i}$ as a function of binding energy is shown
in Fig. \ref{fig:dos}.

However, when $\mathbf{k}$ approaches the $\overline{\Gamma}$ point the
strength of $e$-ph interaction increases significantly. It reaches up to
$1.2-1.4$ in both surface bands. Furthermore the $e$-ph coupling becomes
very sensitive to the energy position of a hole (electron) state in the
surface band. At the same electron momentum, $\lambda_{\mathbf{k}i}$
in the lower surface band can be more than twice as large as in the
higher band.

In the case of Bi(100) surface, the strong energy dependence of
$\lambda_{\mathbf{k}i}$ was explained by simple phase space arguments
with the assumption of a weak energy dependence of the $e$-ph matrix
elements.\cite{gayone03} Here this argument is not valid.  To illustrate
that Fig. \ref{fig:dos} shows the calculated density of electronic states
(DOS). In the surface state energy range,
the density of states is a smooth curve which does not reproduce the
variation of $\lambda_{\mathbf{k}i}$.  Only for hole states with large
momentum in the upper surface band the strength of $e$-ph coupling is
determined to a certain extent by the available phase space.

\begin{figure}
\includegraphics[width=0.8\linewidth]{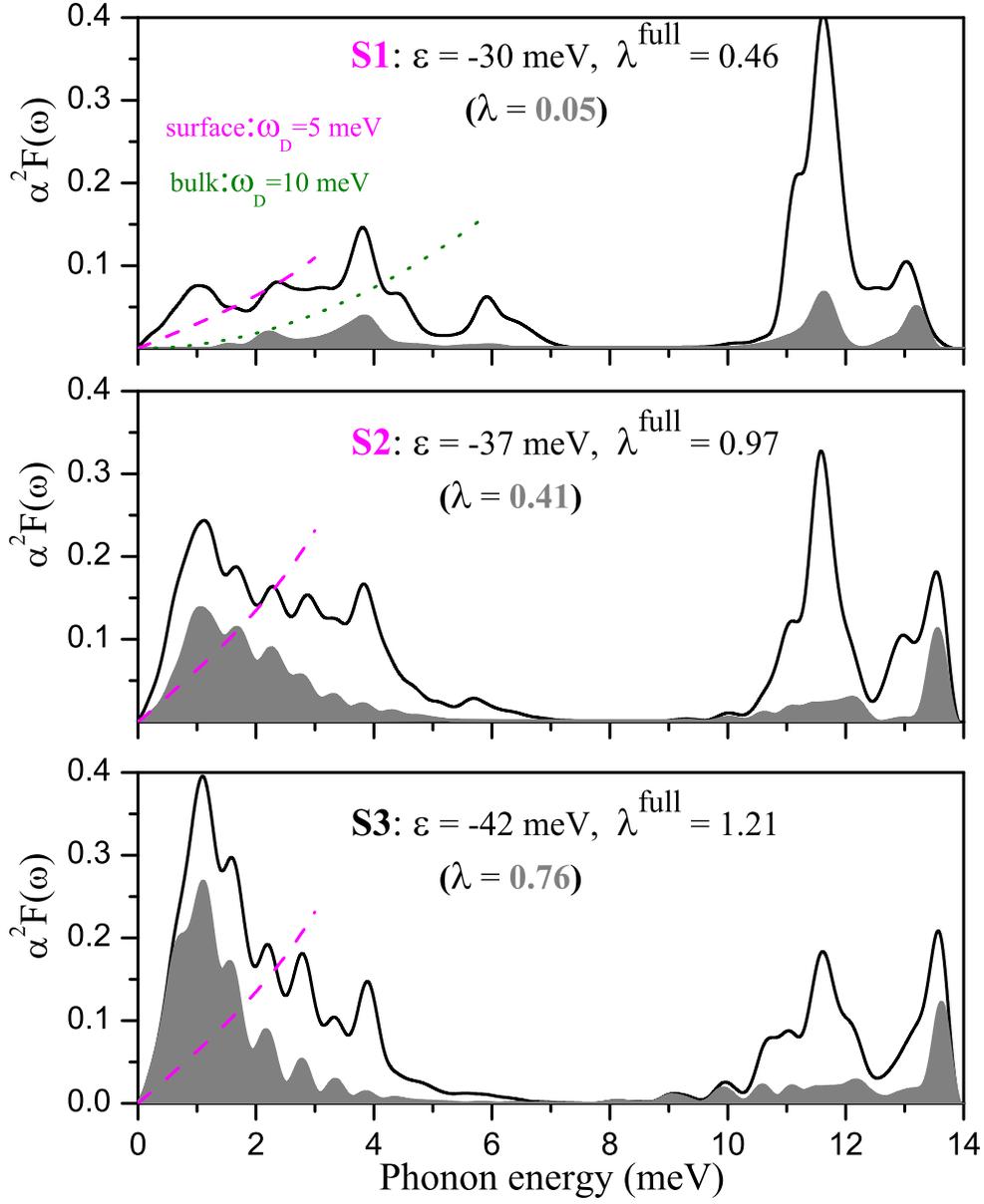}
\caption{(Color Online) Average of the absorption and emission spectral
functions, $\alpha ^{2} F_{\mathbf{k}i}(\omega)$, calculated for surface
electronic states S1, S2, and S3 (see Fig. \ref{fig:lambda}). The
gray areas show the contribution of electronic transitions to the
$\overline{\Gamma}$ point region (Fig. \ref{fig:lambda}).
The dashed (dotted) line indicates the initial slope of a Debye spectrum
assuming a surface (bulk) Debye frequency of 5\,meV (10\,meV) and scaled 
by the corresponding coupling constant.
}\label{fig:alfa}
\end{figure}

The scattering processes leading to such a strong $e$-ph interaction at
small electron momenta can be derived from the corresponding spectral
functions. Fig. \ref{fig:alfa} shows the Eliashberg functions calculated
for surface states S1, S2, and S3 (see Figs. \ref{fig:lambda} and
\ref{fig:dos}), which have similar binding energies but different electron
momenta. Only the average of the emission and adsorption spectral function
is shown because both parts nearly coincide.

A distinctive feature of all $\alpha ^{2} F_{\mathbf{k}i}(\omega)$
shown in Fig.~\ref{fig:alfa} is that the lattice vibrations with small
energies (and small wave vectors) get involved deeply into the scattering
processes of electrons even though their contribution to the phonon
density of states is negligible (Fig.~\ref{fig111pdos}). The type of
transitions they are connected to depends on the momentum position
of the electronic state in the surface bands. In the case of state
S1, the peak in the spectral function at $\sim$1 meV is related to
the nearest intraband transitions. With decreasing electron momentum
(S1$\rightarrow$S2 $\rightarrow$S3) the small energy peak caused by
the intraband scattering remains, but in addition the possibility
of scattering to electronic states near the $\overline{\Gamma}$
point increases. Such transitions via long wavelength phonons with
small energies give a large contribution to the coupling constant
$\lambda$ according to its definition (Eq. \ref{eq:lambda}). The latter
contribution increases rapidly as the electron momentum approaches the
$\overline{\Gamma}$ point and is responsible for the sharp increase of
$\lambda_{\mathbf{k}i}$ at small electron momentum and at energies, where
electronic states around the $\overline{\Gamma}$ point are available for
phonon-mediated transitions.  These states are essentially bulk-like.
Specifically, while the dispersion of the two bands of the surface
states (Fig.\,\ref{fig:lambda}) suggests that they rise above the Fermi
level when approaching $\overline{\Gamma}$, in reality both bands
quickly acquire a bulk-like character, and a true surface localized
state at $\overline{\Gamma}$ appears at much lower energies ($\approx
350$\,meV below $E_F$).  Thus, states near $\overline{\Gamma}$ involved
in the small-momentum transitions are bulk-like.  This low-frequency
contribution clearly increases the coupling strength, but it is also
associated with a larger uncertainty, because for the present 12 bilayer
slab, the bulk-projected part is represented only by a few bands in this
energy region.  Calculations with much thicker slabs would be required for
proper convergence of this contribution, which is currently not feasible.

Attempts to measure the coupling strength of Bi(111) surface states
focused on the part of the hole pocket closest to $\overline{\Gamma}$,
which corresponds approximately to our state S2. From measurements of
the self-energy, Ast and H\"{o}chst extracted coupling constants of
$\lambda = 0.6 \pm 0.05$ and $\lambda = 2.3 \pm 0.2$ when approximating 
the Eliashberg function by a Debye spectrum with either bulk
($\omega_D=10$\,meV) or surface Debye frequencies ($\omega_D=5$\,meV),
respectively.\cite{ast02} Kirkegaard {\em et al.} later argued that
by taking into account the finite spectrometer energy resolution
these values should be corrected down to couplings of the order
0.4.\cite{kirkegaard05} Such a value was obtained by Gayone {\em et al.}
from data on temperature dependent momentum distribution curves near
the Fermi level crossing.\cite{gayone05} This analysis was also based
on a Debye model for the Eliashberg function.

The unusual shape of the calculated spectral functions, in particular
for the S2 and S3 states, renders the applicability of the Debye model
rather questionable. As shown by the dashed lines in Fig.\,\ref{fig:alfa},
the Debye approximation\cite{hellsing02} misses the strong enhancement
of $\alpha ^{2} F_{\mathbf{k}i}(\omega)$ at $\omega \approx 1$\,meV. On
contrast, the Debye model might be more appropriate for an analysis in
the momentum region $k \approx (0.3-0.5) |\overline{\Gamma \rm M|}$,
where this low-energy peak is not very pronounced. For example, the
initial slope of the Eliashberg function for the S1 state is reasonably
well described by the Debye model if the surface Debye frequency of
5\,meV is used (Fig.\,\ref{fig:alfa}).

\section{Summary}
\label{sec_sum}

We have performed a comprehensive investigation of the structural,
electronic, lattice dynamical and electron-phonon coupling properties of
the Bi(111) surface within density functional perturbation theory, taking
into account the spin-orbit coupling consistently. Although some details
of the electronic structure of Bi(111) depend on the slab thickness,
we found that the lattice dynamics of Bi(111) is practically converged
already for slabs of 6 bilayers. Changes in the dynamical couplings are
confined essentially to interatomic bonds in the first two bilayers, where
also the main structural relaxation occurs. The surface phonon spectrum
exhibits super-bulk modes which lie above the bulk spectrum. In addition,
the layer-projected vibrational density of states is enhanced at lower
frequencies for the first bilayer, consistent with the observed enhanced
average vibrational amplitudes of the surface atoms. A calculation of
electronic state dependent coupling to phonons gave moderate coupling
strengths of 0.45 for surface states with larger momenta.
Surface states close to $\overline{\Gamma}$
are predicted to have significantly higher couplings of the order of 1.
This increase is connected
to an enhanced scattering via phonons with long wavelengths and small
energies into bulk-like electronic states near $\overline{\Gamma}$.
The calculated coupling for states of the hole Fermi surface closest
to $\overline{\Gamma}$ is about twice as large as those deduced
from experiment.
This discrepancy may be partly due to insufficient thickness
of the slab, but also may result from an inappropriate assumption in
the experimental analysis.
We found that the state dependent Eliashberg functions strongly deviate in
shape from that of a Debye model, indicating that the use of the latter in an analysis
of electronic self-energies from photoemission spectra is not justified
and likely fails to give correct results for the coupling strength.

\section{Acknowledgments}

This work of MAO and TSR was partially funded by DOE Grant
DE-FG02-07ER46354. 



\clearpage
\begin{thebibliography}{99}

\bibitem[$^\ddagger$]{corr} Corresponding author
\bibitem{shoenberg39} D. Shoenberg, Proc. Roy. Soc. A. Math. and Phys. Sci. {\bf 170}, 341 (1939).
\bibitem{weitzel} B. Weitzel and H. Micklitz, Phys. Rev. Lett. {\bf 66}, 385 (1991).
\bibitem{gayone03} J.~E. Gayone, S.~V. Hoffmann, Z.~Li, and Ph. Hofmann, Phys. Rev. Lett. {\bf 91}, 127601 (2003).
\bibitem{huang13} G.~Q.~Huang and J.~Yang, J. Phys.: Condens. Matter {\bf 25}, 175004 (2013).
\bibitem{tian08} M. Tian, N. Kumar, M. H. W. Chan, and T. E. Mallouk, Phys. Rev. B {\bf 78}, 045417 (2008).
\bibitem{chen71} T. T. Chen, J. D. Leslie, and H. J. T. Smith, Physica {\bf 55}, 439 (1971).
\bibitem{heid10} R. Heid, K. P. Bohnen, I. Y. Sklyadneva, and E. V. Chulkov, Phys. Rev. B {\bf 81}, 174527 (2010).
\bibitem{hofmann06} Ph. Hofmann, Prog. Surf. Sci. {\bf 81}, 191 (2006).
\bibitem{jezequel86} G. Jezequel, Y. Petroff, R. Pinchaux, and F. Yndurain, Phys. Rev. B {\bf 33}, 4352 (1986).
\bibitem{hengsberger00} M. Hengsberger, P. Segovia, M. Garnier, D. Purdie, and Y. Baer, 
                        Eur. Phys. J. B {\bf 17}, 603 (2000).
\bibitem{ast01} Ch.~R. Ast and H. H{\"o}chst, Phys. Rev. Lett. {\bf 87}, 177602 (2001).
\bibitem{ast02} Ch.~R. Ast and H. H{\"o}chst, Phys. Rev. B {\bf 66}, 125103 (2002).
\bibitem{ast04} Ch.~R. Ast and H. H{\"o}chst, Phys. Rev. B {\bf 70}, 245122 (2004).
\bibitem{koroteev04} Yu.~M. Koroteev, G.~Bihlmayer, J.~E. Gayone, E.~V. Chulkov, S.~Bl{\"u}gel, 
                     P.~M. Echenique, and Ph. Hofmann, Phys. Rev. Lett. {\bf 93}, 046403 (2004).
\bibitem{kirkegaard05} C.~Kirkegaard, T.~K. Kim, and Ph. Hofmann, New J. Phys. {\bf 7}, 99 (2005).
\bibitem{gayone05} J.~E. Gayone, C.~Kirkegaard, J.~W. Wells, S.~V. Hoffmann, Z.~Li, and Ph. Hofmann, 
                   Appl. Phys. A {\bf 80}, 943 (2005).
\bibitem{hofmann09} Ph. Hofmann, I.Yu. Sklyadneva, E.D.L. Rienks, and E.V. Chulkov, 
                    New J. Phys. {\bf 11}, 125005 (2009).
\bibitem{sklyadneva11} I. Y. Sklyadneva, G. Benedek, E. V. Chulkov, P. M. Echenique, R. Heid, K. P. Bohnen, 
                       and J. P. Toennies, Phys. Rev. Lett. {\bf 107}, 095502 (2011).
\bibitem{tamtoegl10} A. Tamt\"{o}gl, M. Mayrhofer-Reinhartshuber, N. Balak, W. E. Ernst, and K. H. Rieder, 
                     J. Phys.: Condens. Matter {\bf 22}, 304019 (2010).
\bibitem{tamtoegl13} A. Tamt\"{o}gl, P. Kraus, M. Mayrhofer-Reinhartshuber, D. Campi, M. Bernasconi, G. Benedek, 
                     and W. E. Ernst, Phys. Rev. B {\bf 87}, 035410 (2013).
\bibitem{koroteev08} Y. M. Koroteev, G. Bihlmayer, E. V. Chulkov, and S. Bl\"{u}gel, Phys. Rev. B {\bf 77}, 045428 (2008).
\bibitem{murray07} \'{E}. D. Murray, S. Fahy, D. Prendergast, T. Ogitsu, D. M. Fritz, and D. A. Reis, 
                   Phys. Rev. B {\bf 75}, 184301 (2007).
\bibitem{diaz07} L. E. D\'{i}az-S\'{a}nchez, A. H. Romero, and X. Gonze, Phys. Rev. B {\bf 76}, 104302 (2007).
\bibitem{chis13} V.~Chis, G.~Benedek, P.~M.~Echenique, and E.~V.~Chulkov, Phys. Rev. B {\bf 87}, 075412 (2013).
\bibitem{yang13} J.~Yang, G.~Q.~Huang, and X.~F.~Zhu, Phys. Stat. Solidi B {\bf 250}, 1937 (2013).
\bibitem{benedek14} G. Benedek, M. Bernasconi, K.-P. Bohnen, D. Campi, E. V. Chulkov, P. M. Echenique, R. Heid, 
                  I. Yu. Sklyadneva, and J. P. Toennies, Phys. Chem. Chem. Phys. {\bf 16}, 7159 (2014).
\bibitem{hedin71} L. Hedin and B. I. Lundqvist, J. Phys. C {\bf 4}, 2064 (1971).
\bibitem{perdew96} J. P. Perdew, K. Burke, and M. Ernzerhof, Phys. Rev. Lett. {\bf 77}, 3865 (1996).
\bibitem{vande85} D. Vanderbilt, Phys. Rev. B {\bf 32}, 8412 (1985).
\bibitem{louie79} S. G. Louie, K. M. Ho, and M. L. Cohen, Phys. Rev. B {\bf 19}, 1774 (1979).
\bibitem{meyer} B. Meyer, C. Els\"{a}sser, M. F\"{a}hnle, FORTRAN90 Program
                  for Mixed-Basis Pseudopotential Calculations for Crystals,
                  Max-Planck-Institut f\"ur Metallforschung, Stuttgart
                  (unpublished).
\bibitem{kleinman80} L. Kleinman, Phys. Rev. B {\bf 21}, 2630 (1980).
\bibitem{press92} W. H. Press, S. A. Teukolsky, W. T. Vetterling, and B. P. Flannery, Numerical Recipes in Fortran,
                The Art of Scientific Computing, vol. 2nd Ed. (Cambridge Cambridge University Press,1992).
\bibitem{zein84} N.~E. Zein, Fiz. Tverd. Tela (Leningrad) {\bf 26}, 3028 (1984) [Sov. Phys. Solid State {\bf 26}, 1825 (1984)].
\bibitem{baroni87} S. Baroni, P. Giannozzi, and A. Testa, Phys. Rev. Lett. {\bf 58}, 1861 (1987).
\bibitem{heid99} R. Heid and K. P. Bohnen, Phys. Rev. B {\bf 60}, R3709 (1999).
\bibitem{gianozzi91} P. Giannozzi, S. de Gironcoli, P. Pavone, and S. Baroni, Phys. Rev. B {\bf 43}, 7231 (1991).
\bibitem{heid03} R. Heid and K. P. Bohnen, Phys. Rep. {\bf 387}, 151 (2003).
\bibitem{dresselhaus71} M. S. Dresselhaus, The physics of semimetals and narrow-gap semiconductors: 
                Proceedings (Supplement No. 1 to the Journal of physics and chemistry of solids, 
                v. 32, Pergamon Press, 1971).
\bibitem{schiferl68} D. Schiferl and C. S. Barrett, J. Appl. Cryst. {\bf 2}, 30 (1969).
\bibitem{farlane71} R. E. MacFarlane, The physics of semimetals and narrow-gap semiconductors: 
                    Proceedings (Supplement No. 1 to the Journal of physics and chemistry of solids, 
                    v. 32, Pergamon Press, 1971).
\bibitem{hoehne77} J. H\"{o}hne, U. Wenning, H. Schultz, and S. H\"{u}fner, Z. Physik B {\bf 27}, 297 (1977).
\bibitem{lee14} S. Lee, K. Esfarjani, T. Luo, J. Zhou, Z. Tian, and G. Chen, Nat. Commun. {\bf 5}, 3525 (2014).
\bibitem{jona67} F. Jona, Surf. Sci. {\bf 8}, 57 (1967).
\bibitem{moenig05} H. M\"{o}nig, J. Sun, Y. M. Koroteev, G. Bihlmayer, J. Wells, E. V. Chulkov, K. Pohl, 
                   and P. Hofmann, Phys. Rev. B {\bf 72}, 085410 (2005).
\bibitem{hirahara06} T. Hirahara, T. Nagao, I. Matsuda, G. Bihlmayer, E.V. Chulkov, Yu.M. Koroteev, 
                     P. M. Echenique, M. Saito, and S. Hasegawa, Phys. Rev. Lett. {\bf 97}, 146803 (2006).
\bibitem{alcantara} M.~Alc\'{a}ntara Ortigoza {\em et al.}, to be published.
\bibitem{ramanathan55} K.G. Ramanathan and T.M. Srinivasan, Phys. Rev. {\bf 99}, 442 (1955).
\bibitem{grimvall81} G.~Grimvall, {\it The Electron--Phonon Interaction in Metals}
                   (North-Holland, New York, 1981).
\bibitem{hellsing02} we used the 2D$_2$ model of: B. Hellsing, A. Eiguren, and E V Chulkov, J. Phys.: Condens. Matter {\bf 14}, 5959 (2002).

\end{thebibliography}
\end{document}